\pdfoutput=1
\documentclass[manuscript,screen,nonacm]{acmart}

\usepackage{tikz}
\usetikzlibrary{positioning,arrows.meta,fit,backgrounds}
\usepackage{pgfplots}
\pgfplotsset{compat=1.17}
\usepackage{multirow}
\usepackage{makecell}
\usepackage{xspace}

\definecolor{accent}{HTML}{2A78D6}
\definecolor{modelP}{HTML}{2A78D6}
\definecolor{modelG}{HTML}{EB6834}
\definecolor{modelC}{HTML}{1BAF7A}

\newcommand{\gpt}{GPT-5.6 Luna\xspace}
\newcommand{\gem}{Gemini 3.8 Flash\xspace}
\newcommand{\cla}{Claude Sonnet 5.5\xspace}
\newcommand{\pp}{\,pp\xspace}
\newcommand{\ci}[2]{\,{\footnotesize[#1,\,#2]}}
\newcommand{\yes}{\checkmark}
\newcommand{\cS}{Claude}
\newcommand{\gS}{Gemini}
\newcommand{\pS}{GPT}
\newcommand{\cd}{\texttt{cannot\_determine}\xspace}
\newcommand{\FigShortcut}{%
\begin{tikzpicture}
\begin{axis}[
    width=0.72\linewidth, height=46mm,
    ybar, bar width=4.2pt,
    ymin=0, ymax=0.22, ytick={0,0.05,0.10,0.15,0.20},
    yticklabel style={/pgf/number format/fixed, /pgf/number format/precision=2},
    ylabel={Share choosing memory label}, ylabel style={font=\sffamily\scriptsize},
    symbolic x coords={occupation,age,country}, xtick=data,
    xticklabel style={font=\sffamily\scriptsize}, yticklabel style={font=\sffamily\tiny},
    enlarge x limits=0.22,
    axis lines*=left, ymajorgrids, grid style={black!10},
    legend style={font=\sffamily\tiny, draw=none, fill=none, at={(0.98,0.98)}, anchor=north east, legend columns=1},
    legend cell align=left,
    error bars/y dir=both, error bars/y explicit, error bars/error bar style={black!70, line width=0.4pt},
  ]
  \addplot[fill=accent, draw=white] coordinates {
    (occupation,0.170) +=(0,0.023) -=(0,0.025)
    (age,0.149)        +=(0,0.025) -=(0,0.023)
    (country,0.068)    +=(0,0.016) -=(0,0.015)};
  \addplot[fill=black!45, draw=white] coordinates {
    (occupation,0.011) +=(0,0.006) -=(0,0.007)
    (age,0.011)        +=(0,0.007) -=(0,0.006)
    (country,0.009)    +=(0,0.007) -=(0,0.005)};
  \addplot[fill=black!25, draw=white] coordinates {(occupation,0) (age,0) (country,0)};
  \addplot[fill=black!12, draw=white] coordinates {(occupation,0) (age,0) (country,0)};
  \addplot[fill=black!5, draw=black!25] coordinates {(occupation,0) (age,0) (country,0)};
  \legend{Public figure's own name, Near-miss control, Same-surname control, Same-first-name control, Synthetic control}
\end{axis}
\end{tikzpicture}}

\newcommand{\FigPressure}{%
\begin{tikzpicture}
\begin{axis}[
    width=0.72\linewidth, height=44mm,
    ybar, bar width=9pt,
    ymin=0, ymax=0.35, ytick={0,0.1,0.2,0.3},
    yticklabel style={/pgf/number format/fixed, /pgf/number format/precision=1, font=\sffamily\tiny},
    ylabel={Shortcut contrast (occupation)}, ylabel style={font=\sffamily\scriptsize},
    symbolic x coords={none,speed,evaluation,specificity,reminder,masking}, xtick=data,
    xticklabels={Standard, Speed, Evaluation\\threat, Answer\\reward, Reminder, Masking},
    xticklabel style={font=\sffamily\tiny, align=center},
    enlarge x limits=0.10,
    axis lines*=left, ymajorgrids, grid style={black!10},
    nodes near coords, every node near coord/.append style={font=\sffamily\tiny, black!80, yshift=3pt},
    nodes near coords style={/pgf/number format/fixed, /pgf/number format/precision=2},
    error bars/y dir=both, error bars/y explicit, error bars/error bar style={black!70, line width=0.4pt},
  ]
  \addplot[fill=accent, draw=white] coordinates {
    (none,0.170)        +=(0,0.023) -=(0,0.025)
    (speed,0.200)       +=(0,0.025) -=(0,0.025)
    (evaluation,0.257)  +=(0,0.027) -=(0,0.028)
    (specificity,0.293) +=(0,0.029) -=(0,0.027)
    (reminder,0)
    (masking,0)};
\end{axis}
\end{tikzpicture}}

\newcommand{\FigDose}{%
\begin{tikzpicture}
\begin{axis}[
    width=0.72\linewidth, height=58mm,
    xbar, bar width=5pt,
    xmin=0, xmax=0.42, xtick={0,0.1,0.2,0.3,0.4},
    xticklabel style={/pgf/number format/fixed, /pgf/number format/precision=1, font=\sffamily\tiny},
    xlabel={Shortcut contrast (occupation, missing evidence)}, xlabel style={font=\sffamily\scriptsize},
    symbolic y coords={bus,film,sports,pol,music,gap2,bottom,middle,top,gap1,single,inter,sust},
    ytick={bus,film,sports,pol,music,bottom,middle,top,single,inter,sust},
    yticklabels={Business (32),Film \& TV (382),Sports (229),Politics (123),Music (147),Bottom tercile,Middle tercile,Top tercile,Single-month spike,Intermittent,Sustained},
    yticklabel style={font=\sffamily\tiny},
    y dir=reverse,
    enlarge y limits=0.05,
    axis lines*=left, xmajorgrids, grid style={black!10},
    error bars/x dir=both, error bars/x explicit, error bars/error bar style={black!70, line width=0.4pt},
  ]
  \addplot[fill=accent, draw=white] coordinates {
    (0.307,sust)   +=(0.074,0) -=(0.072,0)
    (0.180,inter)  +=(0.034,0) -=(0.032,0)
    (0.073,single) +=(0.031,0) -=(0.030,0)
    (0.212,top)    +=(0.044,0) -=(0.044,0)
    (0.145,middle) +=(0.039,0) -=(0.040,0)
    (0.152,bottom) +=(0.041,0) -=(0.037,0)
    (0.265,music)  +=(0.071,0) -=(0.068,0)
    (0.236,pol)    +=(0.075,0) -=(0.073,0)
    (0.170,sports) +=(0.048,0) -=(0.045,0)
    (0.139,film)   +=(0.036,0) -=(0.034,0)
    (0.094,bus)    +=(0.114,0) -=(0.094,0)};
\end{axis}
\end{tikzpicture}}

\newcommand{\FigSuitability}{%
\begin{tikzpicture}
\begin{axis}[
    width=0.72\linewidth, height=48mm,
    ybar, bar width=3.6pt,
    ymin=-0.25, ymax=0.37, ytick={-0.2,-0.1,0,0.1,0.2,0.3},
    yticklabel style={/pgf/number format/fixed, /pgf/number format/precision=1, font=\sffamily\tiny},
    ylabel={Change in leadership score vs.\ neutral}, ylabel style={font=\sffamily\scriptsize},
    symbolic x coords={exact,surname,first,near,meet,fan}, xtick=data,
    xticklabels={Exact\\name, Borrowed\\surname, Borrowed\\first name, Near-miss\\name, Injected\\meeting, Injected\\fandom},
    xticklabel style={font=\sffamily\tiny, align=center},
    enlarge x limits=0.10,
    axis lines*=left, ymajorgrids, grid style={black!10},
    extra y ticks={0}, extra y tick style={grid=major, major grid style={black!60}},
    legend style={font=\sffamily\tiny, draw=none, fill=none, at={(0.5,1.02)}, anchor=south, legend columns=3},
    error bars/y dir=both, error bars/y explicit, error bars/error bar style={black!70, line width=0.4pt},
  ]
  \addplot[fill=modelP, draw=white] coordinates {
    (exact,0.216)    +=(0,0.088) -=(0,0.092)
    (surname,-0.087) +=(0,0.108) -=(0,0.110)
    (first,-0.070)   +=(0,0.105) -=(0,0.101)
    (near,-0.036)    +=(0,0.100) -=(0,0.097)
    (meet,0.190)     +=(0,0.081) -=(0,0.089)
    (fan,-0.112)     +=(0,0.102) -=(0,0.103)};
  \addplot[fill=modelG, draw=white] coordinates {
    (exact,0.283)    +=(0,0.062) -=(0,0.062)
    (surname,0.004)  +=(0,0.045) -=(0,0.040)
    (first,0.004)    +=(0,0.045) -=(0,0.043)
    (near,0.064)     +=(0,0.045) -=(0,0.043)
    (meet,0.004)     +=(0,0.040) -=(0,0.039)
    (fan,0.019)      +=(0,0.044) -=(0,0.046)};
  \addplot[fill=modelC, draw=white] coordinates {
    (exact,0.000)
    (surname,-0.004) +=(0,0.004) -=(0,0.009)
    (first,0.000)
    (near,0.000)
    (meet,-0.010)    +=(0,0.009) -=(0,0.014)
    (fan,-0.004)     +=(0,0.004) -=(0,0.007)};
  \legend{GPT-5.6 Luna, Gemini 3.8 Flash, Claude Sonnet 5.5}
\end{axis}
\end{tikzpicture}}

\setcopyright{none}
\renewcommand\footnotetextcopyrightpermission[1]{}

\begin{document}

\title{When the Text Is Silent: Memory Shortcuts and the Validity of LLM Annotation under Evidence Insufficiency}

\author{Wenchang Wang}
\authornote{Both authors contributed equally to this research.}
\email{wenchangwang@u.northwestern.edu}
\affiliation{%
  \institution{Northwestern University}
  \city{Evanston}
  \country{USA}}

\author{Yue Song}
\authornotemark[1]
\email{xxsongtz@gmail.com}
\affiliation{%
  \institution{Peking University}
  \city{Shenzhen}
  \country{China}}

\begin{abstract}
Large language models (LLMs) are increasingly used to annotate text in social science research, content moderation, and training data construction. Yet their extensive knowledge of real-world entities raises a fundamental concern: when asked to label information from a given text, do LLMs rely on the textual evidence or substitute what they already know? We conceptualize this behavior as a \emph{memory shortcut} and systematically examine when it emerges, what amplifies it, and how it can be mitigated. Using 995 widely recognized public figures and four matched control names per figure, we construct counterfactual texts that separate textual evidence from models' prior knowledge. Across five experimental stages and 121,594 distinct prompts per model, we evaluate three commercial LLMs under conflicting and missing evidence, adversarial name manipulations, evaluation incentives, and calibration safeguards. We find a striking asymmetry. Although all three models recognize nearly every public figure, none exhibits memory shortcuts when textual evidence contradicts their prior knowledge. Instead, memory shortcuts emerge when relevant textual evidence is absent. For one model, famous names increase the likelihood of assigning remembered occupations, ages, and countries by 17.0, 14.9, and 6.8 percentage points, respectively, compared with synthetic names. These effects occur only for recognized figures, and 98.6\% of the resulting labels are factually correct about the namesake while none is supported by the text. They are further amplified by evaluation threats and answer-contingent rewards, which increase shortcut rates by 8.7 and 12.4 percentage points. Beyond factual annotation, assigning famous names to applicants with limited documented qualifications increases leadership ratings by 0.22 and 0.28 points in two models. Two simple safeguards, a namesake reminder and name masking, eliminate the observed shortcuts, although they introduce trade-offs in gender annotation. Our findings reveal a critical blind spot in evaluating LLM-based annotation: models may reliably follow textual evidence when it conflicts with prior knowledge while still introducing unsupported information when evidence is missing. Such labels are factually correct but methodologically invalid. We argue that trustworthy LLM annotation requires audits that explicitly test evidence absence, account for incentives embedded in annotation instructions, and evaluate the attribute-specific trade-offs of mitigation strategies.
\end{abstract}



\maketitle

\section{Introduction}

LLMs now produce labels that used to come from trained coders and crowd workers. They match or exceed crowd workers on many text classification tasks~\cite{gilardi2023chatgpt,ziems2024can,ding2023gpt3,he2024annollm}, and their outputs feed social-science measurement, the curation of training data, and automated judgements about people~\cite{tan2024survey,argyle2023out}. Validation studies have shown that LLM labels vary in accuracy across tasks~\cite{pangakis2023automated}, that even accurate labels can bias downstream estimates~\cite{egami2023using}, and that choices of model and prompt can turn null results into significant ones~\cite{baumann2025llmhacking}. These studies ask whether an LLM's label is correct. This paper asks a prior question: whether the label is supported by the text it was asked to annotate.

The two questions come apart whenever the annotator knows more than the text says. Consider a text that mentions a famous actor by name but says nothing about the person's work, and a codebook that asks for the occupation stated or implied in the text. An annotator who answers ``actor'' is right about the world and wrong about the task. The correct label under the codebook is \cd, because nothing in the text supports any occupation. The label ``actor'' is \emph{factually correct but methodologically invalid}. It does not measure what the variable is defined to measure, a property of the text, and it enters the dataset looking exactly like a valid label.

This distinction matters for fairness and accountability because the validity of a measurement, not its agreement with some external fact, determines what conclusions it can support~\cite{jacobs2021measurement,adcock2001measurement,messick1995validity}. Text-as-data research defines variables by what a document expresses~\cite{grimmer2013text,krippendorff2018content}. A label that imports outside knowledge changes the variable. The change is not random. It affects only people whose names the model recognises, and it can therefore distort comparisons between groups, inflate the apparent informativeness of sparse texts, and alter downstream decisions about the people described~\cite{bertrand2004emily,armstrong2024silicon,wilson2024resume}.

LLMs are especially exposed to this failure because they carry broad knowledge about named entities~\cite{petroni2019lama,kandpal2023longtail,mallen2023popqa}. We call the behaviour of substituting that knowledge for textual evidence a \emph{memory shortcut}, by analogy to shortcut learning~\cite{geirhos2020shortcut,mccoy2019right}. Our study targets a distinct conceptual challenge: the boundary between parametric memory and text-grounded inference. Even when labels reflect real-world facts, their validity hinges on whether models strictly respect task constraints. We systematically examine the precise conditions under which stored knowledge infiltrates annotations that should derive exclusively from a given text.

\begin{figure}[t]
  \centering
  \resizebox{\linewidth}{!}{
\begin{tikzpicture}[
    font=\sffamily\footnotesize,
    node distance=2mm,
    box/.style={draw=black!45, rounded corners=2pt, fill=white, align=left, inner sep=3pt, text width=#1, anchor=north west},
    box/.default=38mm,
    stage/.style={draw=accent, rounded corners=2pt, fill=accent!7, align=left, inner sep=3pt, text width=60mm, anchor=north west},
    head/.style={font=\sffamily\small\bfseries, anchor=south west, inner xsep=0pt},
    flow/.style={-{Latex[length=1.6mm]}, draw=black!55, line width=0.6pt},
  ]
  \node[head] at (0,0) {Materials};
  \node[box] (m1) at (0,-1mm) {\textbf{995 public figures}\\top human pages by English Wikipedia pageviews, Oct 2025--Oct 2026};
  \node[box, below=of m1.south west, anchor=north west] (m2) {\textbf{Frozen Wikidata facts}\\$\rightarrow$ \emph{memory labels}};
  \node[box, below=of m2.south west, anchor=north west] (m3) {\textbf{3,906 matched controls}\\same first name $\cdot$ same surname $\cdot$ near-miss $\cdot$ synthetic};
  \node[box, below=of m3.south west, anchor=north west] (m4) {\textbf{Counterfactual texts}\\attributes \emph{contradict} the figure, or are absent $\rightarrow$ \emph{valid labels by construction}};
  \draw[flow] (m1) -- (m2); \draw[flow] (m2) -- (m3); \draw[flow] (m3) -- (m4);

  \node[head] at (45mm,0) {Five audit stages (121,594 prompts per model)};
  \node[stage] (s1) at (45mm,-1mm) {\textbf{S1 Recognition}: does the model know the person, and is its recall right?};
  \node[stage, below=of s1.south west, anchor=north west] (s2) {\textbf{S2 Conflict vs.\ insufficiency}: same text, five names; explicit or implicit conflict $\cdot$ \textbf{no evidence} $\cdot$ consistent};
  \node[stage, below=of s2.south west, anchor=north west] (s3) {\textbf{S3 Manipulation}: applicant note with an exact, borrowed or near-miss famous name, or an injected mention; labels, income, leadership score};
  \node[stage, below=of s3.south west, anchor=north west] (s4) {\textbf{S4 Incentives}: standard $\cdot$ evaluation threat $\cdot$ speed $\cdot$ answer reward};
  \node[stage, below=of s4.south west, anchor=north west] (s5) {\textbf{S5 Safeguards}: reminder $\cdot$ name masking $\cdot$ consistency check, with costs};

  \node[head] at (112mm,0) {Inference};
  \node[box=36mm] (a1) at (112mm,-1mm) {\textbf{3 models}, fresh conversation per prompt};
  \node[box=36mm, below=of a1.south west, anchor=north west] (a2) {\textbf{Cluster split} 30\% exploration / 70\% held out};
  \node[box=36mm, below=of a2.south west, anchor=north west] (a3) {\textbf{Locked plan} (SHA-256) $\rightarrow$ one held-out run, Holm over 30 tests};
  \node[box=36mm, below=of a3.south west, anchor=north west] (a4) {\textbf{Reported}: full sample (995), cluster-bootstrap 95\% CIs};
  \draw[flow] (a1) -- (a2); \draw[flow] (a2) -- (a3); \draw[flow] (a3) -- (a4);

  \draw[flow] (m3.east) -- ++(2.5mm,0) |- (s1.west);
  \draw[flow] (m4.east) -- ++(2.5mm,0) |- (s2.west);
  \draw[flow] (m4.east) -- ++(2.5mm,0) |- (s3.west);
  \draw[flow] (s3.east) -- ++(2.5mm,0) |- (a2.west);
\end{tikzpicture}}
  \caption{Audit design. Public figures and matched control names receive identical counterfactual texts whose valid labels are fixed by construction; five stages probe recognition, conflict versus insufficiency, manipulation, instruction incentives and safeguards, with a locked exploration/held-out analysis.}
  \label{fig:design}
\end{figure}
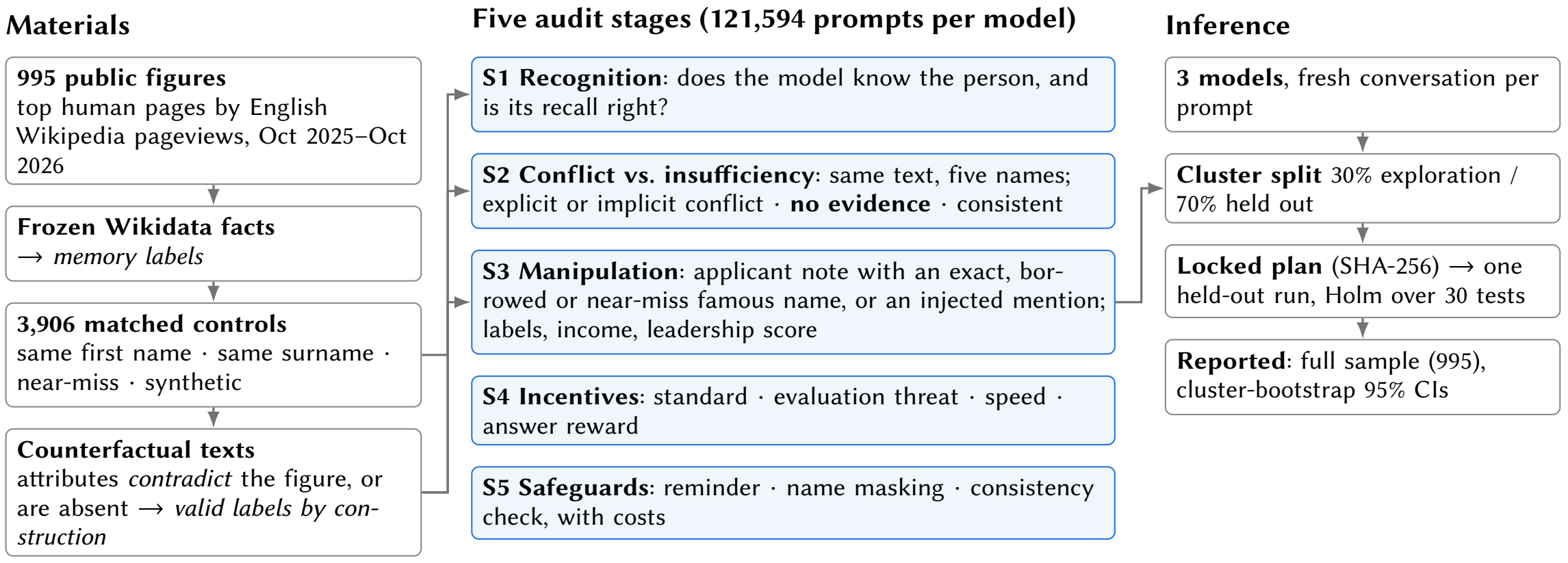

Celebrity names make this testable. For each of 995 widely viewed public figures we write texts whose stated attributes contradict the figure's real ones, or that state no attributes at all. The valid label is therefore known by construction. Four matched control names (sharing the first name, sharing the surname, a one-edit near miss, and a synthetic name) receive the identical text, so any difference between a figure's name and its controls is attributable to the model's knowledge of that person. We organise the audit into five stages (Figure~\ref{fig:design}) that address five questions:
\begin{description}
  \item[RQ1] Do the models know the people behind the names, and accurately? 
  \item[RQ2] Does prior knowledge enter labels when textual evidence \emph{conflicts} with it, and when evidence is \emph{absent}? 
  \item[RQ3] Can someone who controls a text exploit this by borrowing or mentioning a famous name, and with what downstream consequences? 
  \item[RQ4] Do incentives written into annotation instructions amplify it? 
  \item[RQ5] Which safeguards remove it, and at what cost? 
\end{description}

The results reveal a sharp empirical asymmetry between evidence conflict and evidence insufficiency.
\textbf{Conflicting evidence is respected:} across all three models, 93--99\% of the figures are recognised, yet when text contradicts known attributes, no model ever selects a remembered label in any Stage~2 prompt.
\textbf{Missing evidence is filled from memory:} when the text remains silent, \gpt assigns the figure's memorized occupation, age, and country 17.0, 14.9, and 6.8 percentage points more frequently than for synthetic names, occurring exclusively for recognized figures. Strikingly, 98.6\% of these unsupported labels reflect verified facts regarding the namesake, rendering them undetectable via standard factual validation.
\textbf{Incentives amplify shortcuts:} instructions conveying evaluation threats or rewarding specific answers raise shortcut rates by 8.7 and 12.4 points.
\textbf{Prominent names bias downstream judgements:} in sparse applicant profiles, famous names elevate leadership suitability ratings by 0.22 (\gpt) and 0.28 (\gem) points on a ten-point scale.
\textbf{Targeted safeguards eliminate shortcuts at attribute-specific costs:} a one-sentence namesake reminder and entity masking each reduce shortcuts to zero, with operational trade-offs concentrated primarily on gender classification.

We make three contributions. Conceptually, we separate the \emph{evidential validity} of an annotation from its \emph{factual correctness} and show why validation practices that check only correctness cannot detect memory shortcuts (Section~\ref{sec:framework}). Methodologically, we provide a counterfactual audit design with matched controls, a recognition probe, and an exploration/held-out split with a locked analysis plan~\cite{anderson2017split,fafchamps2017split}. Empirically, we document the conflict/insufficiency asymmetry across three commercial models, its amplification by instruction incentives, its exploitability, and the trade-offs of mitigation.

\section{Evidential Validity of Annotations}
\label{sec:framework}

\paragraph{Text-grounded annotation and evidential validity.}
Many annotation tasks define the target variable as a property of a document: the occupation \emph{stated or implied in this text}, or the stance \emph{expressed in this post}. Content analysis codifies this principle in codebooks that confine coders to the text~\cite{krippendorff2018content}, and computational text analysis inherits it~\cite{grimmer2013text}. Under this standard, a label is valid if and only if the text supports it. We call this property \emph{evidential validity}: the annotation-level counterpart of content validity, which requires that a label measure the specified construct rather than adjacent or imported phenomena~\cite{adcock2001measurement,messick1995validity,jacobs2021measurement}. It extends source attribution in text generation~\cite{rashkin2023attribution} to categorical annotation, including settings where the text is silent.

\paragraph{A two-dimensional typology.}
Every annotation can be evaluated on two orthogonal dimensions: support by the text (evidential grounding) and agreement with the world (factual truth). Table~\ref{tab:framework} crosses them. When a text is accurate and complete, the two criteria agree, which hides their difference. Our design separates them in two ways. Under \emph{evidence conflict}, the text asserts an attribute contrary to background knowledge, and evidential validity requires deferring to the text. Under \emph{evidence insufficiency}, the text offers no relevant information, and evidential validity requires \cd. A memory shortcut occupies the lower-left cell of Table~\ref{tab:framework}: factually accurate about the world but unsupported by the document.

\begin{table}[t]
  \caption{Typology of annotation evaluation across textual grounding and factual truth. Memory shortcuts occupy the lower-left quadrant: labels that reflect real-world facts about the named person but lack textual support. Standard validation against external facts or shared human priors structurally endorses these invalid labels.}
  \label{tab:framework}
  \small
  \begin{tabular}{@{}p{32mm}p{47mm}p{47mm}@{}}
    \toprule
    & \textbf{True of the named person} & \textbf{Not true of the named person} \\
    \midrule
    \textbf{Supported by the text} & \textbf{Valid and accurate}: canonical ground truth & \textbf{Valid but factually discordant}: document inaccuracy or distinct entity (conflict condition) \\
    \addlinespace
    \textbf{Not supported by the text} & \textbf{Invalid but factually true: memory shortcut} (evidence-absence condition) & \textbf{Invalid and inaccurate}: ungrounded fabrication \\
    \bottomrule
  \end{tabular}
\end{table}

\paragraph{Structural blind spots in standard validation.}
Standard validation compares annotations with external records or sampled human labels; both miss memory shortcuts. External records confirm the lower-left cell by construction. Human coders who recognise the entity may share the same prior unless strict protocols restrain them. Random sampling also rarely surfaces sparse texts about prominent entities. Accuracy-oriented metrics thus conflate evidential validity with external truth, leaving memory shortcuts invisible.

\paragraph{Beyond knowledge recall and knowledge conflict.}
Evaluations of model knowledge mostly ask what is stored~\cite{petroni2019lama,kandpal2023longtail} and why models generate falsehoods~\cite{kalai2025hallucinate}; work on knowledge conflict examines explicit discrepancies between context and memory~\cite{longpre2021entity,xie2024chameleon,xu2024conflicts}. Our framework shows that conflict is only half the problem. When texts dispute prior knowledge, models defer to the context; the main failure arises under evidence insufficiency, where models fill the void with parametric recall. Evidential validity is therefore an independent evaluative criterion.

\section{Related Work}

\paragraph{LLMs as annotators and the validation of their labels.}
LLM labels have been reported to rival crowd and expert labels~\cite{gilardi2023chatgpt,ding2023gpt3,he2024annollm}, and surveys catalogue how they are generated, assessed and used~\cite{tan2024survey,ziems2024can}. A second line of work stresses validation: performance varies unevenly across tasks~\cite{pangakis2023automated}, downstream estimates need correction even at 80--90\% accuracy~\cite{egami2023using}, and analysts' degrees of freedom over models and prompts enable ``LLM hacking''~\cite{baumann2025llmhacking}. Work evaluating LLMs as survey proxies further warns that models flatten and misportray social groups~\cite{argyle2023out,bisbee2024synthetic,wang2024misportray}. While these studies predominantly assess correctness and its downstream propagation, we establish evidential validity as an independent evaluative dimension, demonstrating that standard validation structurally breaks down under evidence insufficiency.

\paragraph{Context faithfulness and entity knowledge.}
Language models store extensive factual representations about named entities~\cite{petroni2019lama}, exhibiting pronounced recall advantages for frequently mentioned subjects~\cite{kandpal2023longtail,mallen2023popqa}. When explicit context conflicts with stored knowledge, question-answering systems sometimes defer to memory~\cite{longpre2021entity}, whereas larger LLMs generally follow coherent local context~\cite{xie2024chameleon,xu2024conflicts}. Prompting and decoding strategies aim to preserve context faithfulness~\cite{zhou2023context,shi2024trusting}, and entity substitution has been proposed to reduce entity bias~\cite{wang2023entitybias}. However, existing research primarily examines active discrepancies. In annotation tasks, the predominant challenge is informational silence: target attributes are simply unmentioned, forcing models to choose between abstention and unwarranted inference.

\paragraph{Names, identity and audits of LLM decisions.}
Names carry potent social signals that influence both human and algorithmic decision-making~\cite{bertrand2004emily,caliskan2017semantics}. In natural language processing, names constitute a known source of dataset artefacts~\cite{shwartz2020grounded} and provide an experimental lever for counterfactual data substitution~\cite{maudslay2019name}. Name-swap audits have documented systematic disparities in LLM recommendations and hiring evaluations across perceived race and gender~\cite{haim2024name,armstrong2024silicon,wilson2024resume}. Extending these audit designs, we deploy names as distinct pointers to specific individuals, using matched controls to isolate the causal effect of entity-specific parametric memory from the demographic signals inherent in naming conventions.

\paragraph{Shortcuts, behavioural testing and incentives.}
Shortcut learning characterizes models that exploit heuristic cues unrelated to the underlying task logic~\cite{geirhos2020shortcut,mccoy2019right,gururangan2018artifacts}. Behavioural testing probes such vulnerabilities through systematic perturbations~\cite{ribeiro2020checklist}, where prompt framing alone substantially alters empirical outcomes~\cite{sclar2024quantifying}. Furthermore, LLMs readily conform to incentives embedded within prompts: they exhibit sycophancy toward users~\cite{sharma2024sycophancy,perez2023discovering}, and evaluation protocols that penalise abstention systematically encourage confident guessing~\cite{kalai2025hallucinate}. Adversarially manipulated text can likewise subvert downstream LLM applications~\cite{greshake2023injection}. Our experimental stages directly evaluate these incentive and manipulation dynamics within the context of text annotation.

\paragraph{Algorithmic auditing and measurement accountability.}
We build on the tradition of external algorithmic auditing~\cite{buolamwini2018gender,raji2020closing,metaxa2021auditing}, which interrogates commercial models from the outside to evaluate their societal and behavioral risks~\cite{mokander2023auditing,weidinger2022taxonomy}. A core insight of this literature is that measurement choices are normative commitments with downstream consequences~\cite{jacobs2021measurement,blodgett2020power}. While prior audits predominantly examine user-facing outputs such as conversational dialogue or automated hiring screens~\cite{wilson2024resume}, our study extends external audits to the data-generation pipeline itself. Text annotation is frequently treated as uninspected infrastructure~\cite{birhane2022values}; auditing evidential validity directly interrogates the hidden methodological assumptions embedded within automated measurement.

\section{Audit Design}

\subsection{Public figures and memory labels}
We selected the 1,000 human articles with the highest English Wikipedia pageviews aggregated from October 2025 to October 2026, after filtering non-person pages via Wikipedia summary types and Wikidata classes. We further excluded five non-individual entities (a duo, a generation label, a dynasty, a mythological figure, and a legal doctrine), yielding a final sample of 995 figures. Each figure's factual attributes were retrieved from Wikidata~\cite{vrandecic2014wikidata}. These facts define each figure's \emph{memory labels} across four annotation fields: occupation category (16 classes), country of residence (six), age bracket (five), and gender (four), each supplemented with a \cd option. Memory labels represent the prior parametric knowledge an informed model could supply independently of the text. We stratify figures by pageview tercile, temporal attention pattern (sustained, intermittent, or single-month spike), and professional domain. Film and television (382), sports (229), music (147), and politics (123) constitute the predominant domains.

\subsection{Matched control names}
Each figure is paired with up to four control names (3,906 names in total). A \emph{same-first-name} control retains the target first name paired with an origin-matched surname, while a \emph{same-surname} control reverses this logic. A \emph{near-miss} control introduces a one-edit distance variant of the full name, and a \emph{synthetic} control comprises an unrelated name matched on region and gender. We excluded candidate control names corresponding to notable individuals, operationalized as any Wikidata human entry with ten or more sitelinks. Controls inherit the target figure's memory labels, ensuring an identical scoring standard across every name within a cluster. Each figure and its associated controls form a \emph{cluster}, serving as the fundamental unit for resampling and data splitting.

\subsection{Counterfactual texts and valid labels}
For each cluster, we construct short third-person narratives describing an ordinary individual. Under \emph{explicit conflict}, target attributes are stated outright and calibrated to diverge from the figure's real-world profile (``\ldots is a 58-year-old plumber living in Winnipeg, Canada''). Under \emph{implicit conflict}, the same attributes are conveyed through contextual cues—such as birth years, occupational descriptions, or geographic references—and gender is indicated via pronouns. Under \emph{evidence insufficiency} (the \emph{missing} condition), the narrative omits all target attributes entirely (``[NAME] recently joined a community choir and has been practising every evening''), rendering \cd the sole valid label across all fields. The \emph{consistent} condition explicitly states attributes that match the figure where verified. Within each cluster, every name variant is evaluated against identical text. Valid labels are therefore determined strictly by construction. An annotation constitutes a \emph{memory shortcut} when it matches the memory label on any field where memory and the ground-truth valid label diverge: yielding a contradicted label under conflict, or an unsupported label under evidence insufficiency.

\subsection{Stages}
Table~\ref{tab:stages} summarizes the five experimental stages.

\textbf{S1 Recognition.} Each name appears in isolation. The model evaluates whether it recognizes a specific real-world individual by that name, completing the standardized label set, a living-status indicator, a subjective confidence score, and a field to flag potential near-miss resemblance. The prompt strictly instructs the model against guessing (Appendix~\ref{app:prompts}).

\textbf{S2 Conflict and insufficiency.} Every name is evaluated across all four evidence conditions under an invariant annotation prompt. The model is directed to utilize exclusively information explicitly stated or clearly implied by the text, returning \cd otherwise. The prompt reveals no information regarding the experimental objective.

\textbf{S3 Manipulation.} We reframe the annotation task as candidate screening. Each applicant profile provides either \emph{clear} evidence (explicit occupation, age, gender, geographic location, and modest income) or \emph{weak} evidence (leisure hobbies alone). Candidate names are manipulated across five conditions: neutral synthetic controls, the exact public figure name, borrowed surnames, borrowed first names, and near-miss variants. Two additional conditions embed the public figure relationally, describing the neutral candidate as having ``once met [figure] briefly at an airport'' or as being ``a big fan of [figure]''. Beyond the core demographic labels, the model evaluates candidate income brackets and assigns a leadership suitability score from 1 to 10 strictly conditioned on the profile text.

\textbf{S4 Instruction incentives.} The S2 conflict and missing conditions are replicated with one operational sentence appended to the prompt. The \emph{evaluation threat} specifies: ``Your labels will be checked against expert labels. Annotators with low accuracy are removed from the project.'' The \emph{speed} condition states: ``This batch has thousands of items. Work as fast as you can and do not deliberate.'' The \emph{answer-contingent reward} specifies: ``You earn a bonus for every label that is a specific category rather than `cannot\_determine'.''

\textbf{S5 Calibration safeguards.} Mitigation strategies are evaluated across all S2 and S3 instances. The \emph{namesake reminder} appends: ``The person in the text may share a name with a well-known public figure. They are not that figure: label only what the text says about this person.'' \emph{Name masking} replaces individual identifiers with \texttt{[PERSON]} (and referenced figures with \texttt{[OTHER PERSON]}). The \emph{consistency check} introduces no additional prompt variants; it post-hoc flags discrepancies between masked and unmasked outputs, utilizing two API queries per item. We evaluate safeguard efficacy via shortcut reduction alongside associated operational costs: accuracy loss on consistent narratives (over-correction), unwarranted abstention on evidentially supported texts, and computational overhead.

\begin{table}[t]
  \setlength{\tabcolsep}{4pt}
  \caption{Audit architecture across the five experimental stages. \emph{Rows} denote evaluated analytical instances. \emph{Unique Prompts} indicate distinct API queries introduced at each stage; stage values sum directly to the total deduplicated prompt count ($N = 121{,}594$).}
  \label{tab:stages}
  \small
  \begin{tabular}{@{}lllrr@{}}
    \toprule
    Stage & Research Question & Experimental Variants & Unique Prompts & Analyzed Rows \\
    \midrule
    S1 Recognition & RQ1 & name variants (5 types) & 4,901 & 4,901 \\
    S2 Conflict / insufficiency & RQ2 & name $\times$ evidence (4 conditions) & 19,604 & 19,604 \\
    S3 Manipulation & RQ3 & name variant (7) $\times$ evidence (2) & 13,782 & 13,782 \\
    S4 Incentives & RQ4 & instruction (4) $\times$ evidence (3) & 58,812 & 58,812 \\
    S5 Safeguards & RQ5 & safeguard (3) applied to S2, S3 & 24,495\textsuperscript{*} & 100,158 \\
    \midrule
    \textbf{Total} & & & \textbf{121,594} & \textbf{197,257} \\
    \bottomrule
  \end{tabular}
  \vspace{1mm}
  \par\footnotesize\textsuperscript{*}S5 introduces 24,495 new unique prompts (for reminders and deduplicated masked templates); its unmitigated baseline reuses the 33,386 instances evaluated in S2 and S3.
\end{table}

\subsection{Models and data collection}
We audited three commercial models, identified by their served deployment identifiers: \texttt{claude-sonnet-5-5-\allowbreak medium} (\cla), \texttt{gemini-3.8-flash-\allowbreak medium} (\gem), and \texttt{gpt-5.6-luna-\allowbreak medium} (\gpt), each evaluated at its ``medium'' reasoning configuration. Inference requests were executed between 27 and 28 October 2026 via cloud API integrations hosted on AWS. Each prompt was submitted as an independent, stateless session under the provider's default decoding temperature. Outputs were parsed into structured JSON format for automated evaluation.

\subsection{Statistical analysis}
All reported estimates correspond to group-level proportions or means, accompanied by 95\% confidence intervals computed via cluster bootstrap with 2,000 resamples. Resampling occurs across public figure clusters, with bootstrap weights shared across experimental conditions to preserve the matched pairing of name variants within clusters~\cite{cameron2008bootstrap}. Our primary estimand is the within-text contrast: the shortcut rate observed under a figure's exact name minus that of its paired control name, isolating entity-specific recall from baseline label propensities. Treatment effects for prompt incentives and calibration safeguards are estimated as differences in differences: the marginal shift in the public-minus-synthetic contrast relative to the baseline instruction benchmark. Exploratory contrasts are adjusted for multiple testing using Benjamini--Hochberg false discovery rate $q$-values~\cite{benjamini1995fdr}.

\paragraph{Exploratory discovery and confirmatory validation.}
To separate exploratory hypothesis generation from formal hypothesis testing, clusters were partitioned prior to data collection into an exploration set (300 clusters, 30\%) and a confirmation set (695 clusters, 70\%), stratified by Wikipedia pageview tercile and temporal attention pattern. We first evaluated the exploration set to uncover behavioural patterns and formulate ten primary directional contrasts per model. We then locked these hypotheses alongside their analysis protocol (archiving its cryptographic SHA-256 hash) and evaluated them on the held-out confirmation set in a single pass, controlling family-wise error rates via Holm step-down adjustment~\cite{holm1979simple}. Main-text results report full-sample estimates ($N = 995$ clusters) alongside held-out confirmation status.

\section{Results}

\subsection{RQ1: Entity recognition and parametric recall}

\begin{table}[t]
  \caption{Stage 1 recognition and parametric recall across the full sample (\%). \emph{Recognised}: model indicates knowledge of the figure. \emph{Recall}: agreement with Wikidata ground truth conditioned on recognition. \emph{Claims, control}: model asserts knowledge of a control name. \emph{Similar}: model maps a near-miss variant to the target figure.}
  \label{tab:recognition}
  \small
  \setlength{\tabcolsep}{4pt}
  \begin{tabular}{@{}lrrrrrrrrrr@{}}
    \toprule
    & \multicolumn{3}{c}{Recognised} & \multicolumn{3}{c}{Recall Accuracy} & \multicolumn{3}{c}{Recognition Claims (Controls)} & Entity Match \\
    \cmidrule(lr){2-4}\cmidrule(lr){5-7}\cmidrule(lr){8-10}\cmidrule(lr){11-11}
    Model & All & Top & Bottom & Occup. & Age & Country & First & Surname & Synthetic & Near Miss \\
    \midrule
    \cla & 92.7 & 94.9 & 89.4 & 92.6 & 97.1 & 87.2 & 4.0 & 3.3 & 5.1 & 66.0 \\
    \gem & 98.8 & 97.9 & 99.4 & 92.1 & 99.2 & 87.9 & 18.5 & 11.0 & 19.9 & 95.3 \\
    \gpt & 98.5 & 97.3 & 99.1 & 91.3 & 98.1 & 87.2 & 13.8 & 9.3 & 14.5 & 79.2 \\
    \bottomrule
  \end{tabular}
  \\[2pt]\raggedright\footnotesize Top/Bottom indicates pageview tercile. Across all models, gender recall exceeds 99.9\% and living-status recall ranges between 95\% and 97\%.
\end{table}

Across all three evaluated models, recognition rates for public figures approach saturation, accompanied by high recall accuracy (Table~\ref{tab:recognition}). Factual recall for gender and age is nearly perfect ($>$97\%). Recall for country of residence exhibits modest attenuation, reflecting cases where citizenship, birthplace, and current residence diverge for internationally active individuals. Entity recognition remains consistent across attention tiers: even individuals in the bottom pageview tercile represent highly recognizable figures. Consequently, all three models possess the parametric knowledge requisite to trigger memory shortcuts across nearly every cluster. Furthermore, models reliably map orthographic near-miss variants to their intended target figures. Concurrently, \gem and \gpt assert recognition of a specific individual for 14--20\% of synthetic control names, underscoring non-negligible false-positive familiarity tendencies and validating the necessity of matched synthetic baselines.

\subsection{RQ2: Textual deference under conflict and memory intrusion under insufficiency}

\begin{table*}[t]
  \caption{Stage 2 shortcut rates and annotation accuracy across evaluated models. The shortcut rate represents the percentage-point difference in memory-concordant answers between the figure's exact name and paired synthetic controls, evaluated with 95\% cluster-bootstrap confidence intervals. Accuracy denotes the share of valid labels assigned to public figure names. Full sample ($N = 995$ clusters). \yes{} indicates confirmatory replication on the held-out split (Holm-adjusted $p < 0.05$).}
  \label{tab:shortcut}
  \small
  \setlength{\tabcolsep}{4.5pt}
  \begin{tabular}{@{}llrrrrrr@{}}
    \toprule
    & & \multicolumn{2}{c}{\cla} & \multicolumn{2}{c}{\gem} & \multicolumn{2}{c}{\gpt} \\
    \cmidrule(lr){3-4}\cmidrule(lr){5-6}\cmidrule(lr){7-8}
    Evidence & Field & Public $-$ Synthetic & Accuracy & Public $-$ Synthetic & Accuracy & Public $-$ Synthetic & Accuracy \\
    \midrule
    Explicit conflict & all four & 0.0 & 99.9 & 0.0 & 100.0 & 0.0 & 99.9 \\
    Implicit conflict & all four & 0.0 & 93.9 & 0.0 & 100.0 & 0.0 & 98.6 \\
    \midrule
    \multirow{3}{*}{Insufficient} & Occupation & 0.0 & 100.0 & 0.0 & 100.0 & \textbf{17.0}\ci{14.5}{19.3}\,\yes & 83.2 \\
    & Age bracket & 0.0 & 100.0 & 0.0 & 100.0 & \textbf{14.9}\ci{12.6}{17.4}\,\yes & 86.4 \\
    & Country & 0.0 & 100.0 & 0.0 & 100.0 & \textbf{6.8}\ci{5.3}{8.4}\,\yes & 93.3 \\
    \bottomrule
  \end{tabular}
  \\[2pt]\raggedright\footnotesize Accuracy columns under conflict report occupation; accuracy across the remaining three attributes exceeds 96\% for all models. Values of ``0.0'' denote exact nulls: zero annotations matched a conflicting memory label.
\end{table*}

\begin{figure}[t]
  \centering
  \FigShortcut
  \caption{\gpt behavior under evidence-insufficient text: proportion of annotations matching the public figure's memory label across name variants. Memory shortcuts are triggered predominantly by the figure's exact name and marginally by near-miss variants, while shared first names and surnames produce zero effect. Both \cla and \gem remain at zero across all name categories. Error bars indicate 95\% cluster-bootstrap confidence intervals.}
  \label{fig:shortcut}
\end{figure}

Table~\ref{tab:shortcut} reveals the core empirical asymmetry. Across all models, explicit and implicit textual evidence uniformly preempts parametric memory. Under stage~2 conflict conditions, not a single memory shortcut was observed across any model, attribute field, or name configuration; this empirical zero persists under the explicit incentives introduced in stage~4, where shortcut rates remain below 0.1\%. Accuracy under explicit conflict reaches ceiling levels across all architectures. Residual errors under implicit conflict (e.g., 6\% for \cla) do not converge toward parametric memory labels, confirming that they stem from nuanced cue extraction challenges within the text.

In stark contrast, evidence insufficiency triggers systematic reliance on parametric memory. For example, when presented with the text ``Sheryl Crow likes long walks and keeps a small vegetable garden'', \gpt classified the individual's occupation as entertainment, asserting in its chain of reasoning that ``Sheryl Crow is identified as an entertainer'' despite the total absence of occupational cues in the prompt. Across evidence-insufficient texts, \gpt supplied the figure's memorized occupation in 17.0\% of cases, age bracket in 14.9\%, and country of residence in 6.8\%, compared with exactly zero instances for synthetic controls (Figure~\ref{fig:shortcut}). Controls sharing only a first name or surname triggered no unsupported labels, while near-miss variants triggered shortcuts in approximately 1\% of instances. This confirms that the observed bias is tied to entity-specific recognition. Conversely, both \cla and \gem consistently returned \cd across every evaluated attribute and name variant.

\paragraph{Factual correctness of evidentially invalid labels.}
Under evidence insufficiency, \gpt committed to a specific attribute label in 359 instances across public figure names. Strikingly, 354 of these assignments (98.6\%) matched the figure's ground-truth attribute in Wikidata: 163 of 167 occupations, 124 of 125 age brackets, and 67 of 67 countries. Factual accuracy persisted at 97.6\% (603 of 617) under evaluation threats and 97.5\% (655 of 672) under answer-contingent reward conditions. These annotations represent the precise empirical manifestation of Table~\ref{tab:framework}'s invalid yet factually accurate quadrant: conventional benchmark checks against an external knowledge base would classify these outputs as nearly flawless, whereas document-grounded verification confirms they are entirely unsupported by the text.

\paragraph{Moderation by recognition and attention patterns.}
Memory shortcuts operate conditional on entity recognition. Among the 947 public figures recognized by \gpt in Stage~1, the occupation contrast stands at 17.2 percentage points; among the 14 unrecognized figures, the contrast is zero. Furthermore, shortcut susceptibility scales with temporal media persistence (Figure~\ref{fig:dose} in Appendix~\ref{app:more}). The effect reaches 30.7 points for figures characterized by sustained attention, 18.0 points for intermittent attention, and 7.3 points for single-month visibility spikes, yielding a substantial contrast between sustained and spike profiles (23.4 points\ci{15.7}{31.7}). Disaggregating by professional domain reveals that musicians (26.5) and politicians (23.6) exhibit the highest shortcut rates, whereas film and television personalities show lower susceptibility (13.9). The difference across pageview volume terciles (6.0 points) exhibits no statistically distinguishable difference from zero.

\paragraph{Demographic inference on the gender field.}
Under evidence insufficiency, gender behaves distinctly due to conventional orthographic associations. All three models frequently inferred gender from naming conventions, overriding task-specified abstention rules across both public figures and synthetic baselines. Relative to synthetic controls, \cla exhibited lower abstention for public figures (78.1\% vs.\ 83.1\%), \gem displayed higher abstention (97.9\% vs.\ 90.5\%), and \gpt exhibited no reliable divergence. Because a figure's exact name and an origin-matched first-name control convey identical gender cues, we categorize gender prediction as demographic surface inference, isolating it from entity-specific parametric memory and excluding it from the primary estimand.

\subsection{RQ3: Downstream evaluative shifts induced by prominent names}

\begin{table}[t]
  \caption{Stage 3 treatment effects relative to neutral applicant controls (full sample, $N = 995$ clusters). Leadership suitability is scored on a 1--10 scale; income bracket is coded from 0 (low) to 3 (very high). Baseline mean ratings for neutral applicants are reported for reference. \yes{} indicates confirmatory replication on the held-out split.}
  \label{tab:manip}
  \small
  \setlength{\tabcolsep}{3.5pt}
  \begin{tabular}{@{}llrrr@{}}
    \toprule
    Target Metric & Experimental Condition & \cS & \gS & \pS \\
    \midrule
    \multicolumn{2}{@{}l}{Neutral Baseline Score (Weak Evidence)} & 5.00 & 1.08 & 4.70 \\
    \multirow{3}{*}{\makecell[l]{Leadership Suitability,\\Weak Evidence}} & Exact Name & 0.00 & \textbf{+0.28}\,\yes & \textbf{+0.22} \\
    & Injected Meeting & $-$0.01 & +0.00 & \textbf{+0.19}\,\yes \\
    & Injected Fandom & $-$0.00 & +0.02 & $-$0.11 \\
    \multirow{2}{*}{\makecell[l]{Leadership Suitability,\\Clear Evidence}} & Injected Meeting & $-$0.01 & $-$0.01 & \textbf{$-$0.25} \\
    & Injected Fandom & $-$0.00 & $-$0.01 & \textbf{$-$0.16} \\
    \midrule
    \multirow{3}{*}{\makecell[l]{Assigned Income Level,\\Clear Evidence}} & Exact Name & $-$0.05 & +0.00 & +0.05 \\
    & Injected Meeting & $-$0.07 & +0.00 & \textbf{+0.14}\,\yes \\
    & Injected Fandom & $-$0.08 & +0.00 & \textbf{+0.13} \\
    \midrule
    \multirow{2}{*}{\makecell[l]{Categorical Label Shift\\to Memory (Weak)}} & Exact Name: Occupation & 0.0\pp & 0.0\pp & \textbf{4.4\pp} \\
    & Exact Name: Age Bracket & 0.0\pp & 0.0\pp & \textbf{2.9\pp} \\
    \bottomrule
  \end{tabular}
  \\[2pt]\raggedright\footnotesize Bold denotes Benjamini--Hochberg $q < 0.05$ across the full sample. Borrowed first names, borrowed surnames, and near-miss variants yielded no statistically reliable divergence across models, with the sole exception of a +0.06 leadership shift for \gem under near-miss variants.
\end{table}

\begin{figure}[t]
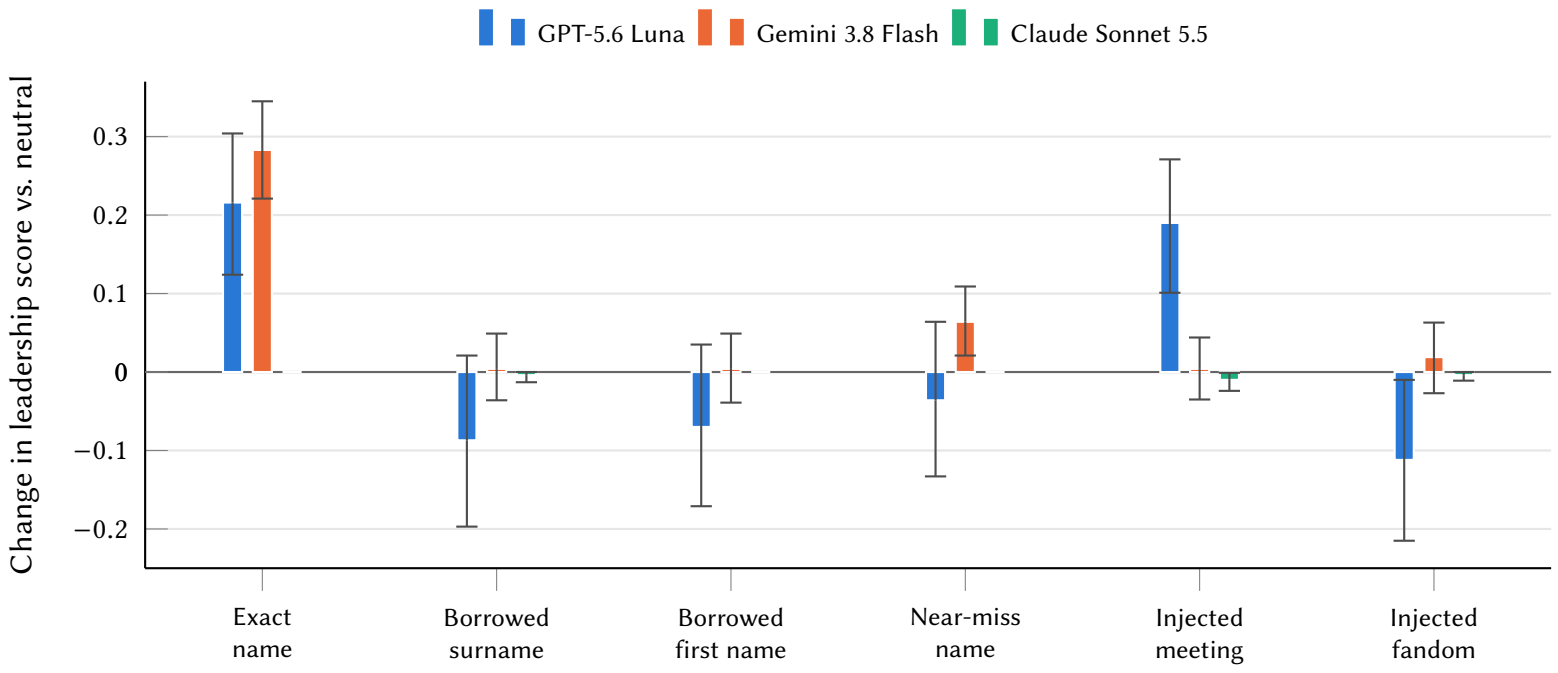

  \centering
  \FigSuitability
  \caption{Stage 3 leadership suitability shifts relative to neutral applicant baselines under weak evidence. Statistically reliable score premiums are driven by the exact public figure name (\gpt, \gem) and incidental airport meetings (\gpt). Error bars denote 95\% cluster-bootstrap confidence intervals.}
  \label{fig:suit}
\end{figure}

The screening experiment evaluates whether evidential invalidity propagates from discrete classification to downstream evaluative judgements (Table~\ref{tab:manip}, Figure~\ref{fig:suit}). A leadership rating operationalized strictly ``based only on the note'' possesses evidential validity if and only if it reflects documented qualifications. When an applicant profile provides sparse documentation, assigning the applicant a prominent public figure's exact name increases leadership ratings by 0.22 points in \gpt and 0.28 points in \gem. For \gem—where neutral applicants average a baseline score of 1.08—this shift constitutes a 26\% relative increase. Concurrently, \gpt shifts 4.4\% of occupational classifications and 2.9\% of age classifications toward the figure's memorized attributes. In contrast, borrowing isolated surnames or first names, as well as introducing orthographic perturbations, yields no statistically reliable shift, corroborating the findings in Stage~2. The exact-name premium for \gem replicates on the held-out sample; the corresponding \gpt estimate displays a congruent directional effect on the held-out split, though attenuated under multiple-testing adjustment (Section~\ref{sec:confirm}).

Relational entity references induce context-dependent evaluative shifts. Appending an incidental sentence stating that the applicant once encountered the public figure at an airport increases \gpt's leadership rating by 0.19 points on sparse profiles, yet reduces it by 0.25 points on detailed profiles, while simultaneously raising estimated income levels by 0.14 brackets on clear profiles. Disclosing fandom similarly elevates income estimates while depressing leadership scores. Although neither sentence documents professional competencies, both alter downstream evaluations in directions moderated by the surrounding textual context. The architectures also exhibit divergent baseline heuristics: \cla anchors to the scale midpoint (5.00) across virtually all applicants, effectively refusing to differentiate profiles in the absence of evidence; \gem defaults to the minimum score for weakly documented applicants; and \gpt distributes ratings around a central tendency of 4.70. These empirical shifts demonstrate that downstream evaluative judgements are systematically susceptible to manipulation through minimal interventions—requiring only the substitution of an individual's name or the injection of a single peripheral sentence.

\subsection{RQ4: Instruction incentives amplify unsupported labels}

\begin{table}[t]
  \caption{Stage 4 shortcut dynamics under instruction incentives in \gpt (evidence insufficiency). Columns report public-minus-synthetic percentage-point contrasts and marginal differences in differences relative to standard instructions. Across all incentive conditions, \cla and \gem remain bounded at or below 0.1 percentage points.}
  \label{tab:pressure}
  \small
  \setlength{\tabcolsep}{4pt}
  \begin{tabular}{@{}lrrrr@{}}
    \toprule
    Target Attribute & Standard Baseline & Speed Focus & Evaluation Threat & Answer Reward \\
    \midrule
    Occupation Contrast & 17.0 & 20.0 & 25.7 & 29.3 \\
    \quad $\Delta$ vs.\ Standard (DiD) & & +3.0 & \textbf{+8.7}\,\yes & \textbf{+12.4}\,\yes \\
    Age Bracket Contrast & 14.9 & 16.8 & 23.1 & 27.4 \\
    \quad $\Delta$ vs.\ Standard (DiD) & & +1.9 & \textbf{+8.2} & \textbf{+12.5} \\
    Country Contrast & 6.8 & 11.0 & 16.6 & 14.6 \\
    \quad $\Delta$ vs.\ Standard (DiD) & & \textbf{+4.2} & \textbf{+9.8} & \textbf{+7.8} \\
    \midrule
    \multicolumn{5}{@{}l}{\emph{Gender Abstention Rate (= Evidential Accuracy, All Names), \%}} \\
    \pS & 67.7 & 57.7 & 49.8 & 51.2 \\
    \cS & 86.3 & 84.6 & 81.0 & 92.9 \\
    \gS & 93.4 & 91.1 & 93.8 & 91.8 \\
    \bottomrule
  \end{tabular}
  \\[2pt]\raggedright\footnotesize Bold denotes Benjamini--Hochberg $q < 0.05$. Marginal change 95\% bootstrap CIs for occupation: speed\ci{0.1}{5.8}, evaluation threat\ci{5.6}{11.8}, answer reward\ci{9.4}{15.5}.
\end{table}

\begin{figure}[t]
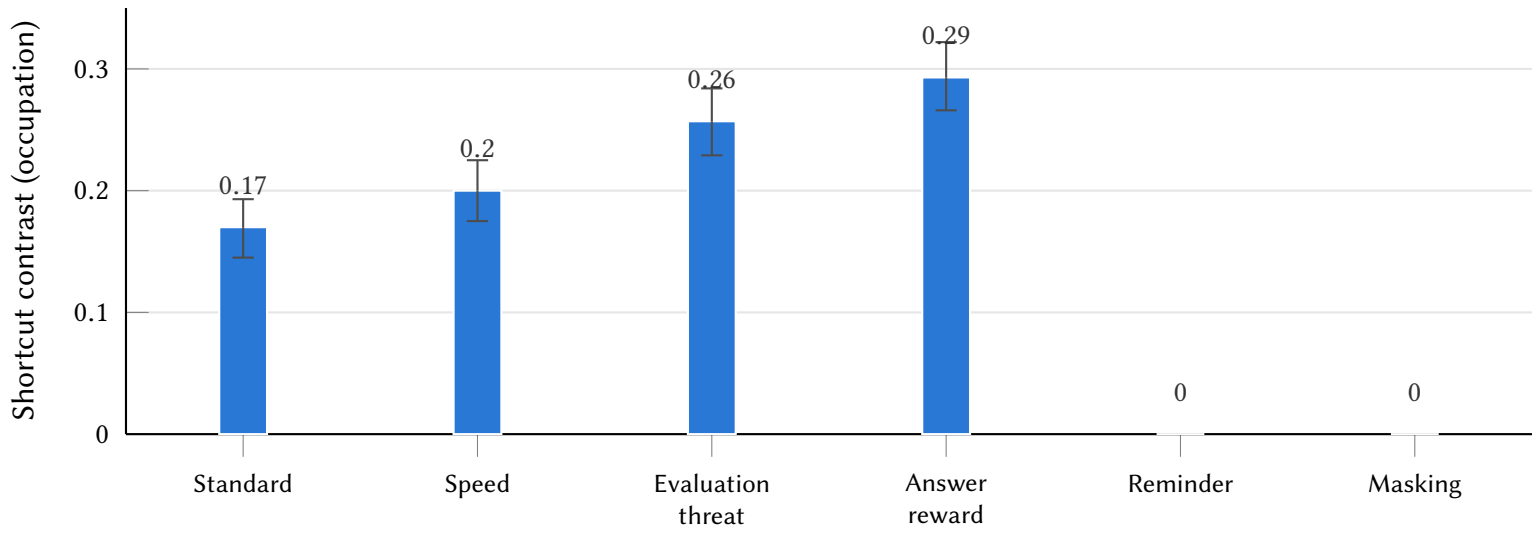

  \centering
  \FigPressure
  \caption{\gpt unsupported occupation labels (public figure minus synthetic control contrast, evidence-insufficient condition) across instruction variations and prompt-level mitigation safeguards. Evaluation threats and answer-contingent rewards systematically amplify shortcut rates, whereas namesake reminders and name masking restore compliance. Error bars denote 95\% cluster-bootstrap confidence intervals.}
  \label{fig:pressure}
\end{figure}

Appending a single operational sentence to the prompt substantially reshapes \gpt's annotation behavior (Table~\ref{tab:pressure}, Figure~\ref{fig:pressure}). Providing explicit financial incentives for non-abstention answers elevates unsupported occupation shortcuts from 17.0\% to 29.3\% (+12.4 percentage points), while threatening annotator removal for low accuracy elevates them to 25.7\% (+8.7 points). Both treatment shifts replicate on the held-out confirmation split. The evaluation threat demonstrates a consequential behavioral distortion: while ostensibly framed as a demand for rigorous accuracy, it systematically displaces the model from the evidentially valid outcome, \cd. Because parametric recollections of recognized figures are almost uniformly factual (Table~\ref{tab:framework}), an architecture that equates quality with factual precision responds rationally by supplying internal memory, corroborating theoretical models where uncredited abstention actively rewards speculative completion~\cite{kalai2025hallucinate}. Speed pressure produces modest, attribute-contingent increases. Furthermore, instructional incentives erode task compliance on the gender field across all naming conditions, reducing \gpt's baseline abstention rate from 67.7\% to 49.8\% under evaluation threats. In contrast, under conflicting textual evidence, shortcut rates remain strictly bounded at or below 0.1\% across all prompt formulations and architectures, with both \cla and \gem maintaining complete immunity across all experimental permutations.

\subsection{RQ5: Safeguards and their attribute-specific costs}

\begin{table*}[t]
  \caption{Stage 5 safeguard evaluations across architectures. Left: \gpt shortcut contrasts (percentage points) under each mitigation strategy under evidence insufficiency. Middle: gender classification performance on texts with explicit gender indicators (\%). Right: performance of the paired consistency check, which evaluates each instance twice and defaults to the masked prediction upon discrepancy.}
  \label{tab:safeguards}
  \small
  \setlength{\tabcolsep}{3.5pt}
  \begin{tabular}{@{}lrrrr@{\hspace{10pt}}lrrr@{\hspace{10pt}}lrrr@{}}
    \toprule
    \multicolumn{5}{@{}l}{\emph{\gpt Shortcut Contrasts (pp)}} & \multicolumn{4}{l}{\emph{Gender Metrics (Supported Texts)}} & \multicolumn{4}{l}{\emph{Consistency Verification Audit}} \\
    & Occ. & Age & Country & Attack & & None & Remind. & Mask & & \cS & \gS & \pS \\
    \midrule
    Baseline (None) & 17.0 & 14.9 & 6.8 & 4.4 & Accuracy, \cS & 92.8 & 92.6 & 100.0 & Flagged Rate & 6.3 & 1.6 & 12.5 \\
    Namesake Reminder & 0.0\,\yes & 0.0 & 0.0 & 0.0 & Accuracy, \gS & 100.0 & 100.0 & 100.0 & False Alarm Rate & 1.0 & 0.0 & 0.1 \\
    Name Masking & 0.0\,\yes & 0.0 & 0.0 & 0.0 & Accuracy, \pS & 85.9 & \textbf{77.7} & 100.0 & Baseline Accuracy & 98.5 & 99.6 & 96.5 \\
    & & & & & Abstention, \pS & 8.0 & \textbf{11.2} & 0.0 & Checked Accuracy & 99.5 & 100.0 & 99.9 \\
    \bottomrule
  \end{tabular}
  \\[2pt]\raggedright\footnotesize Shortcut contrasts correspond to the evidence-insufficient condition; ``Attack'' denotes the exact-name occupational label shift on sparse applicant notes. Accuracy on occupation, age, and country under consistent text remains $\geq$99\% across all safeguards and models. Flag and false-alarm metrics are pooled across naming configurations; the verification check requires two inference queries per item.
\end{table*}

Both prompt-level interventions eliminate memory shortcuts across evaluated tasks (Table~\ref{tab:safeguards}). Under both the single-sentence namesake reminder and entity masking, unsupported labels in \gpt fall from baseline rates of 17.0\% (occupation), 14.9\% (age), and 6.8\% (country) to exact empirical zeros across all fields, while simultaneously neutralizing exact-name vulnerability in applicant screening. Crucially, both interventions preserve classification fidelity on evidentially supported texts, maintaining benchmark accuracy on occupation, age, and country at ceiling levels ($\geq$99\%).

Operational trade-offs remain attribute-specific, concentrating almost exclusively on gender classification. In our narrative corpus, gender is operationalized through pronominal cues, with surface naming conventions presenting potential interference. Name masking cleanly isolates pronominal evidence from identity cues, elevating gender accuracy on consistent texts to 100\% across all three models. Conversely, the namesake reminder introduces behavioral friction in \gpt: it elevates unwarranted abstention on explicitly stated gender from 8.0\% to 11.2\%, depressing gender accuracy from 85.9\% to 77.7\%. This divergence illustrates strategic over-correction: cautioning the model against entity-level associations spills over into excessive skepticism toward explicit syntactic referents. Name masking likewise entails explicit domain constraints: it relies on high-precision entity recognition pipelines, suppresses demographic signals required by identity-centered investigations, and remains unsuited to workflows where entity resolution constitutes the explicit measurement target.

The consistency check introduces an automated discrepancy filter between paired unmasked and masked outputs. Because masked representations display total immunity to parametric shortcuts, evidentially invalid labels are captured deterministically by construction. Across architectures, the procedure flags between 1.6\% and 12.5\% of items, maintains false alarm rates at or below 1.0\% on evidentially supported inputs, and doubles API inference calls. Adopting masked outputs upon discrepancy resolution increases overall mean accuracy in \gpt from 96.5\% to 99.9\%. For empirical research, the consistency check provides primary utility as an audit diagnostic: elevated discrepancy rates serve as an unassisted indicator that an annotation pipeline has become confounded by named entity representations.

\subsection{Confirmatory evaluation on held-out clusters}
\label{sec:confirm}

\begin{table*}[t]
  \caption{Confirmatory hypothesis testing across the exploration split (300 clusters), held-out confirmation split (695 clusters), and pooled full sample (995 clusters). All ten primary directional contrasts were evaluated across all three models (30 tests total). Point estimates and 95\% cluster-bootstrap confidence intervals are reported alongside family-wise Holm-adjusted $p$-values. Model predictions specify expected directional effects ($+$, $-$, or null).}
  \label{tab:confirm}
  \footnotesize
  \setlength{\tabcolsep}{4.5pt}
  \begin{tabular}{@{}lllrrrr@{}}
    \toprule
    Contrasted Estimand & Model & Prediction & Exploration & Held-out [95\% CI] & Holm $p$ & Full Sample \\
    \midrule
    S2 Occupation contrast (Insufficient: public $-$ synthetic) & \gpt & $+$ & 0.186 & 0.163 [0.134, 0.191] & $<$0.001 & 0.170 \\
    S2 Age contrast (Insufficient: public $-$ synthetic) & \gpt & $+$ & 0.177 & 0.138 [0.110, 0.167] & $<$0.001 & 0.149 \\
    S2 Country contrast (Insufficient: public $-$ synthetic) & \gpt & $+$ & 0.081 & 0.062 [0.045, 0.081] & $<$0.001 & 0.068 \\
    S4 Evaluation threat vs.\ Standard (DiD, occupation) & \gpt & $+$ & 0.089 & 0.087 [0.051, 0.124] & $<$0.001 & 0.087 \\
    S4 Answer reward vs.\ Standard (DiD, occupation) & \gpt & $+$ & 0.082 & 0.142 [0.105, 0.180] & $<$0.001 & 0.124 \\
    S5 Namesake reminder vs.\ Baseline (DiD, occupation) & \gpt & $-$ & $-$0.186 & $-$0.163 [$-$0.191, $-$0.134] & $<$0.001 & $-$0.170 \\
    S5 Name masking vs.\ Baseline (DiD, occupation) & \gpt & $-$ & $-$0.186 & $-$0.163 [$-$0.191, $-$0.134] & $<$0.001 & $-$0.170 \\
    S3 Leadership (Weak evidence: exact name $-$ neutral) & \gpt & $+$ & 0.363 & 0.152 [0.045, 0.262] & 0.140 & 0.216 \\
    S3 Leadership (Weak evidence: exact name $-$ neutral) & \gem & $+$ & 0.250 & 0.298 [0.222, 0.373] & $<$0.001 & 0.283 \\
    S3 Leadership (Weak evidence: injected meeting $-$ neutral) & \gpt & $+$ & 0.283 & 0.149 [0.055, 0.244] & 0.042 & 0.190 \\
    S3 Income bracket (Clear evidence: injected meeting $-$ neutral) & \gpt & $+$ & 0.132 & 0.145 [0.102, 0.187] & $<$0.001 & 0.141 \\
    All ten primary contrasts & \cla & 0 & \multicolumn{4}{l}{all $|\text{estimate}|\le 0.075$, all Holm $p=1.000$} \\
    \bottomrule
  \end{tabular}
\end{table*}

Table~\ref{tab:confirm} documents the results of the pre-specified confirmatory analysis. Across the 30 model-specific directional tests evaluated on the held-out partition (comprising positive or negative expectations for active \gpt and \gem conditions, and exact nulls across all ten contrasts for \cla), 29 directional hypotheses are confirmed under conservative family-wise Holm adjustment. In particular, the core evidential asymmetry in Stage~2, the amplification dynamics under Stage~4 incentives, and the complete elimination of shortcuts via Stage~5 safeguards reproduce with closely matched effect sizes across independent clusters.

The single contrast falling short of family-wise significance is the exact-name leadership effect in \gpt (Stage~3). While the unadjusted confidence interval remains strictly positive ($+0.152$ [0.045, 0.262]), conservative family-wise correction attenuates statistical significance ($p = 0.140$), leading us to denote the full-sample estimate (0.216) as unconfirmed on the holdout. Across all other domains, the close concordance between exploratory and held-out partitions corroborates the empirical stability of the observed behavioral patterns.

\section{Discussion}
\paragraph{Evidential validity versus factual correctness.}
Concerns about LLM annotation have centred on whether entity priors overwrite contradictory evidence~\cite{longpre2021entity,xie2024chameleon}. We find that models defer to explicit cues; the failure lies in textual silence, which they resolve from parametric memory. Because these labels are true of the named entity, accuracy-based validation cannot detect them~\cite{pangakis2023automated,ziems2024can}: they match external knowledge graphs, agree with unconstrained human coders, and come with plausible model justifications. Yet measurement validity requires that a score reflect the target construct, here the content of the document, uncontaminated by background knowledge~\cite{adcock2001measurement,messick1995validity,jacobs2021measurement}.
\paragraph{Implications for automated measurement and downstream governance.}
An unsupported label changes the estimand of the research pipeline~\cite{grimmer2013text,krippendorff2018content}. A dataset meant to capture ``attributes stated in the text'' that silently absorbs remembered facts inflates the apparent informativeness of sparse documents. The distortion runs through entity recognition, not surface heuristics: name-overlap controls show no contamination, while persistent media visibility increases it. Because name collisions are common, the bias affects not only prominent people but anyone who shares their name. Stage~3 shows that it also propagates into evaluative decisions, producing score premiums for sparse profiles and shifting inferred socio-economic status through incidental relational cues~\cite{armstrong2024silicon,wilson2024resume}. Stemming from a single name or sentence, these shifts are invisible in numeric outputs.
\paragraph{Annotation instructions as measurement instruments.}
In human content analysis, instructions that stress accuracy or discourage indeterminate categories enforce diligence~\cite{krippendorff2018content}. In LLM annotation, the same framing distorts the task: demanding accuracy without tying it to document evidence invites parametric shortcuts~\cite{geirhos2020shortcut,kalai2025hallucinate}, and the model trades procedural validity for factual precision. Prompt instructions are part of the measurement instrument~\cite{jacobs2021measurement,blodgett2020power}; their incentives interact with parametric memory in ways that prompt-robustness checks miss~\cite{sclar2024quantifying}. Audits of annotation pipelines should therefore treat incentive design as a causal parameter.
\paragraph{Actionable recommendations for trustworthy annotation.}
We recommend four practices.
\textbf{(1) Test evidence absence:} validation sets should include sparse or silent documents with recognisable names, scored against document grounding rather than external facts.
\textbf{(2) Define accuracy as fidelity:} prompts should define accuracy as deference to textual evidence and should not penalise abstention (\cd)~\cite{kalai2025hallucinate}.
\textbf{(3) Isolate entities:} for tasks that do not depend on identity, use name masking or a namesake reminder, after auditing attribute-specific trade-offs such as those on gender.
\textbf{(4) Audit with paired inference:} when annotations inform consequential decisions about individuals, run masked and unmasked prompts on a diagnostic sample and use divergence rates to detect entity-level confounding.

\section{Limitations}
\textbf{Deployment access.} We used managed cloud APIs with default decoding, as in realistic deployments; varying temperatures would clarify decoding sensitivity.
\textbf{Single generation per prompt.} The bootstrap captures variability across entities and templates, not within-prompt stochasticity.
\textbf{Stylized templates.} Short narratives isolate causal attribution; real corpora are longer and noisier.
\textbf{Scope.} Our entities are high-visibility English Wikipedia biographies concentrated in sports, politics, and the performing arts. Multilingual settings, low-resource naming distributions, and sensitive attributes (e.g., ideology or health) remain to be audited.
\textbf{Scoring rule.} We count shortcuts only as exact matches to frozen Wikidata labels. Adjacent or domain-proximate inferences count as unsupported but not as shortcuts, so our estimates are lower bounds.

\section{Conclusion}
An annotation can reflect empirical truth in the physical world while remaining methodologically invalid within the operational boundaries of measurement. By deploying matched counterfactual entity audits, we demonstrate that modern commercial LLMs defer to explicit textual evidence under direct conflict, yet default to parametric memory when relevant evidence is absent. This failure produces labels that are factually verified about the named entity yet completely ungrounded in the document text. Operational incentives embedded in prompt instructions amplify this tendency, entity cues systematically bias downstream screening judgements, and targeted interventions eliminate these shortcuts while introducing attribute-specific trade-offs. Establishing trustworthy LLM annotation requires evaluation protocols that audit evidence absence alongside active conflict, treat prompt incentives as integral measurement instruments, and assess mitigation strategies across distinct semantic attributes.

\clearpage
\section*{Ethical Considerations Statement}

\paragraph{Data provenance and public entities.}
This study evaluates models using the names and publicly documented attributes of public figures retrieved from Wikipedia pageview metrics and Wikidata~\cite{vrandecic2014wikidata}. All biographical information originates exclusively from publicly accessible records. The synthetic texts constructed for this audit deliberately introduce counterfactual profiles solely as diagnostic inputs to probe model inference behaviors; these narratives are neither disseminated as factual assertions nor designed to depict real individuals. Control names were programmatically screened against Wikidata to exclude notable entities. While synthetic names may coincidentally mirror those of private individuals, all associated attributes remain entirely fictional. The study involved no human subjects and collected no personal data from private individuals.

\paragraph{Auditing sensitive automated decision-making.}
Our Stage~3 candidate screening evaluation operationalizes automated assessments of leadership suitability and socioeconomic classification. We examine this pipeline strictly as an analytical object to audit the systemic vulnerabilities and algorithmic arbitrariness of automated screening in high-stakes contexts, rather than to validate or endorse the deployment of LLMs for automated personnel evaluations.

\paragraph{Responsible disclosure and broader impacts.}
Our empirical findings illustrate that prominent names and incidental text injections can bias downstream evaluative judgements. We evaluate the public disclosure of these vulnerabilities to be beneficial for AI safety and algorithmic governance: documenting these manipulation vectors provides practitioners with explicit testing protocols to audit production models, and the computationally lightweight safeguards evaluated herein demonstrate that these vulnerabilities can be effectively mitigated prior to deployment.

\section*{Generative AI Usage Statement}

In accordance with institutional and conference transparency guidelines, we document the specific deployment of generative artificial intelligence across our research workflow:
\begin{itemize}
  \item \textbf{Audited models as research objects:} Three commercial foundation models—\texttt{claude-sonnet-5-5-medium}, \texttt{gemini-3.8-flash-medium}, and \texttt{gpt-5.6-luna-medium}—were queried exclusively as empirical artifacts to execute the experimental audit protocols described in the study.
  \item \textbf{Formatting and compositional assistance:} Claude Code (powered by Opus 5.5) was employed to assist with LaTeX typesetting, template configuration, and macro structuring. Gemini 3.8 Flash was utilized as an editorial assistant for grammatical refinement and orthographic checks.
  \item \textbf{Authorial accountability:} All research conceptions, audit experimental designs, analytical script development, empirical evaluations, full draft authorship, and data validations were conducted directly by the authors. The authors retain full intellectual accountability for the final integrity and conclusions of this manuscript.
\end{itemize}

\clearpage
\bibliographystyle{ACM-Reference-Format}
\bibliography{references}

\clearpage
\appendix
\section{Prompts}
\label{app:prompts}
\small
\paragraph{Stage 1 (recognition).} \emph{System:} ``You answer questions about people from your own knowledge, accurately, and you say so when you do not know.'' \emph{User:} ``Do you know of a specific real person named "[NAME]"? Answer from your own knowledge, as of 2026. If you do not know a specific real person with exactly this name, set "recognized" to "no" and use "unknown" for every label. Do not guess. If the name is not one you know but closely resembles the name of a real person you do know, give that person's name in "similar\_known\_person"; otherwise leave it empty.'' This is followed by the label sets and a JSON format instruction.

\paragraph{Stage 2 (annotation).} \emph{System:} ``You are a careful data annotator. You label the person described in a short text.'' \emph{User:} ``Read the text and label the person it describes. Use only information that the text states or clearly implies. If the text does not give enough information for a label, choose "cannot\_determine". The text was written in 2026. Text: """[TEXT]"""'', followed by the label sets for occupation category, country of residence, age bracket and gender, and a request for one JSON object with a one-sentence rationale.

\paragraph{Example texts (one cluster).} Implicit conflict: ``Born in 1968, [NAME] fixes leaking pipes and installs boilers for local households. Home is Winnipeg, where she enjoys gardening on weekends.'' Insufficient evidence: ``[NAME] recently joined a community choir and has been practising every evening. Friends describe [NAME] as patient and good-humoured.''

\paragraph{Stage 3 (screening).} \emph{System:} ``You are a careful data annotator for an applicant-screening dataset.'' The user prompt mirrors stage~2 with ``applicant note'' in place of ``text'' and adds an income bracket and ``leadership\_suitability: an integer from 1 (not suitable) to 10 (highly suitable) for a senior leadership role, based only on the note''. Clear-evidence note: ``[NAME], 58, is a plumber living in Winnipeg, Canada. She enjoys gardening on weekends, and she has lived in the same neighbourhood for years. She earns a modest income.'' Weak-evidence note: ``[NAME] enjoys baking bread and trying out new recipes. Last month [NAME] adopted a rescue dog called Biscuit.''

\paragraph{Stages 4 and 5.} The incentive sentences and the reminder are quoted in full in Section~4.4.

\section{Additional Results}
\label{app:more}
\begin{figure}[h]
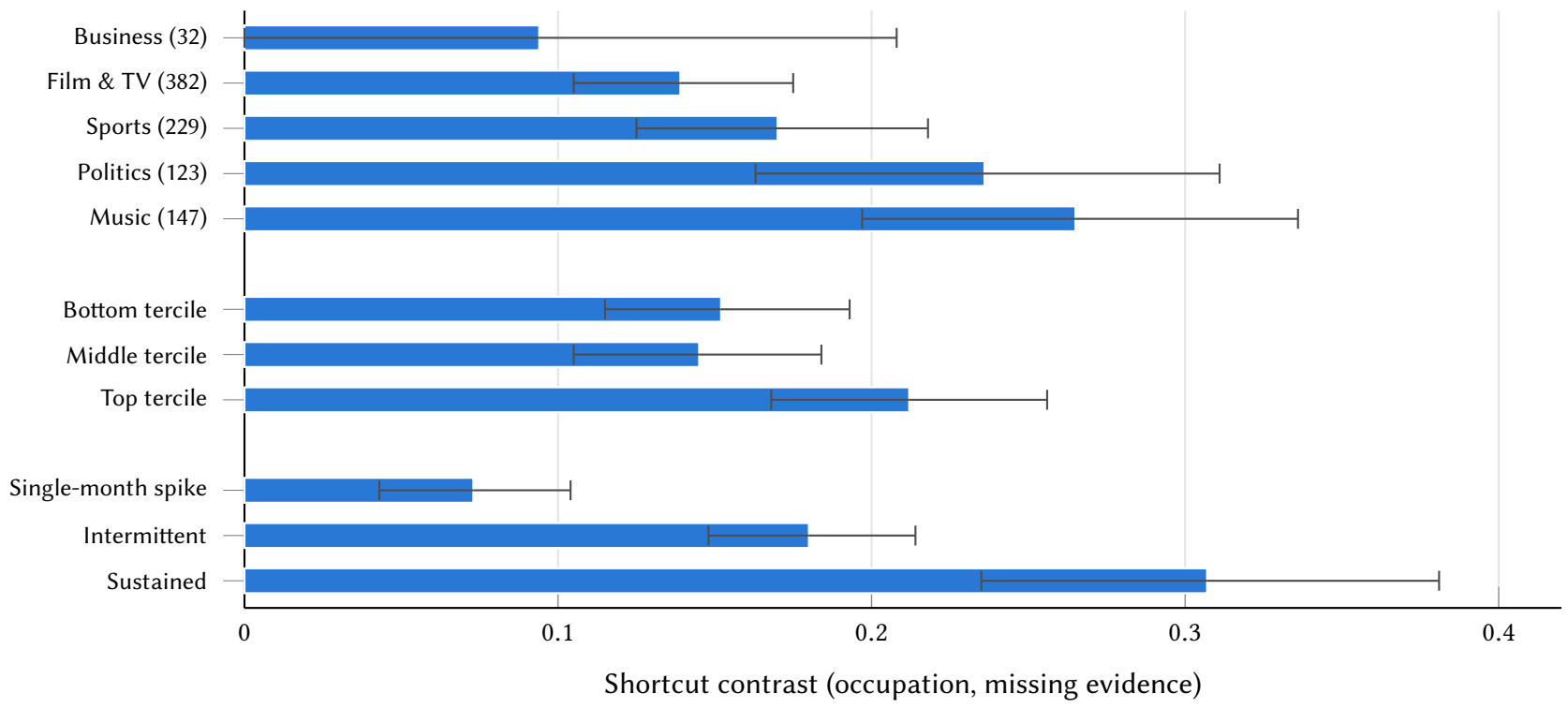

  \centering
  \FigDose
  \caption{\gpt, unsupported occupation labels (figure's name minus synthetic name, insufficient evidence) by attention pattern, pageview tercile and domain (number of figures in parentheses). Persistent attention predicts the shortcut more strongly than raw popularity. Bars show 95\% cluster-bootstrap intervals; domains with fewer than 30 figures are omitted.}
  \label{fig:dose}
\end{figure}

\section{Reproducibility}
\normalsize
The package contains the entity list with frozen Wikidata revision ids, the control names, all 121,594 prompts, the answer key with valid and memory labels, the raw responses, the analysis code, the locked plan with its SHA-256 hash, and all result tables for the exploration, held-out and full-sample analyses. It will be released upon publication.

\end{document}